\documentclass[
	a4paper, 
	10pt, 
	unnumberedsections, 
	twoside, 
]{LTJournalArticle}

\usepackage{cite} 
\usepackage{comment}
\usepackage{stfloats}

\runninghead{} 

\footertext{\textit{RISC-V Summit Europe, Paris, 12-15th May 2025}} 

\title{CVA6S+: A Superscalar RISC-V Core with High-Throughput Memory Architecture} 

\author{%
	Riccardo Tedeschi\textsuperscript{1}\thanks{Corresponding author: \href{mailto:riccardo.tedeschi6@unibo.it}{\tt riccardo.tedeschi6@unibo.it}\\
    This work has received funding from the Swiss State Secretariat for Education, Research, and Innovation (SERI) under the SwissChips initiative.},
    Gianmarco Ottavi\textsuperscript{1},
    Côme Allart\textsuperscript{2,3},
    Nils Wistoff\textsuperscript{4},
    Zexin Fu\textsuperscript{4},\\
    Filippo Grillotti\textsuperscript{5},
    Fabio De Ambroggi\textsuperscript{5},
    Elio Guidetti\textsuperscript{5},
    Jean-Baptiste Rigaud\textsuperscript{3},
    Olivier Potin\textsuperscript{3},\\
    Jean Roch Coulon\textsuperscript{2},
    César Fuguet\textsuperscript{6},
    Luca Benini\textsuperscript{1,4}
    and Davide Rossi\textsuperscript{1}
}

\date{\footnotesize
\textsuperscript{\textbf{1}}DEI, University of Bologna, Bologna, Italy\\
\textsuperscript{\textbf{2}}Thales DIS, Meyreuil, France\\
\textsuperscript{\textbf{3}}Mines Saint-Etienne, CEA, Leti, Centre CMP, F-13541 Gardanne, France\\
\textsuperscript{\textbf{4}}IIS, ETH Zurich, Zurich, Switzerland\\
\textsuperscript{\textbf{5}}STMicroelectronics, Agrate Brianza, Italy\\
\textsuperscript{\textbf{6}}Univ. Grenoble Alpes, Inria, CNRS, Grenoble INP, TIMA, 38000 Grenoble, France}

\renewcommand{\maketitlehookd}{%
	\begin{abstract}
		\noindent Open-source RISC-V cores are increasingly adopted in high-end embedded domains such as automotive, where maximizing instructions per cycle (IPC) is becoming critical. Building on the industry-supported open-source CVA6 core and its superscalar variant, CVA6S, we introduce CVA6S+, an enhanced version incorporating improved branch prediction, register renaming and enhanced operand forwarding. These optimizations enable CVA6S+ to achieve a 43.5\% performance improvement over the scalar configuration and 10.9\% over CVA6S, with an area overhead of just 9.30\% over the scalar core (CVA6). Furthermore, we integrate CVA6S+ with the OpenHW Core-V High-Performance L1 Dcache (HPDCache) and report a 74.1\% bandwidth improvement over the legacy CVA6 cache subsystem.
	\end{abstract}
}

\begin{document}

\maketitle 


\section{Introduction}
CVA6 \cite{zaruba2019cost} is an application-class RISC-V core from the PULP platform, currently maintained by OpenHW Group. It features a six-stage, in-order pipeline and supports both ASIC and FPGA implementations, with a configurable 32- or 64-bit data path width. However, its IPC (Instructions Per Clock) is constrained by its simple, scalar in-order front-end microarchitecture.


Allart et al.~\cite{allart2024using} introduced a superscalar variant, CVA6S, supporting dual-issue execution. CVA6S retains the CVA6 pipeline structure with a 64-bit instruction fetch bus capable of fetching two 32-bit or four 16-bit compressed instructions per cycle. The decode and issue logic are duplicated, and a second ALU is added. We focus on enhancing and extending CVA6S. We introduce the CVA6S+ dual issue RISC-V core, which achieves a 43\% IPC improvement over the scalar CVA6 and an 11\% improvement over CVA6S in the Embench IoT benchmarks.

Additionally, we integrate CVA6S+ with the HPDCache \cite{fuguet2023hpdcache}, a highly configurable, pipelined data cache designed to deliver an Out-of-Order (OoO), non-blocking cache subsystem for RISC-V cores. Compared to the previous cache subsystem, we demonstrate an average 74\% bandwidth improvement in memory-intensive accesses under both regular and irregular access patterns, using the RaiderSTREAM benchmarks. The total area overhead of CVA6S+ over the scalar CVA6 core is less than 10\%.
\section{Implementation}
Building on CVA6S, we introduce key enhancements to improve its microarchitecture to boost performance:
\begin{enumerate}
    \item \textbf{Register Renaming:} To mitigate Write-After-Write (WAW), register renaming tracks the latest instruction writing to integer or floating-point registers, ensuring correct operand forwarding.

    \item \textbf{Branch Prediction:} A two-level branch predictor with private history per entry (128 entries, 3-bit history) is integrated, reducing misprediction penalties by 30\% across the Embench-IoT suite compared to the simpler bimodal predictor used in CVA6 and CVA6S.

    \item \textbf{ALU-to-ALU Operand Forwarding:} To reduce execution latency, lightweight operand forwarding is introduced for cases where two ALU instructions are issued in the same cycle. This allows the second instruction to directly use the result of the first one without delay.

    \item \textbf{FPU Integration:} The floating-point unit (FPU) is integrated into the dual-issue CVA6S+, as CVA6S lacks this feature. To minimize area overhead, it shares the write-back (WB) port with the secondary ALU, requiring additional hazard logic to prevent contention.
\end{enumerate}

Furthermore, we replace the blocking architecture of the legacy WB cache with the HPDcache. It features a three stage pipeline with multiple request ports, a deep MSHR, out-of-order execution, a hardware prefetcher, and support for write-through and write-back policies, cache management, and atomic operations.


\begin{figure*} [!t]
    \centering
    \includegraphics[width=1\linewidth]{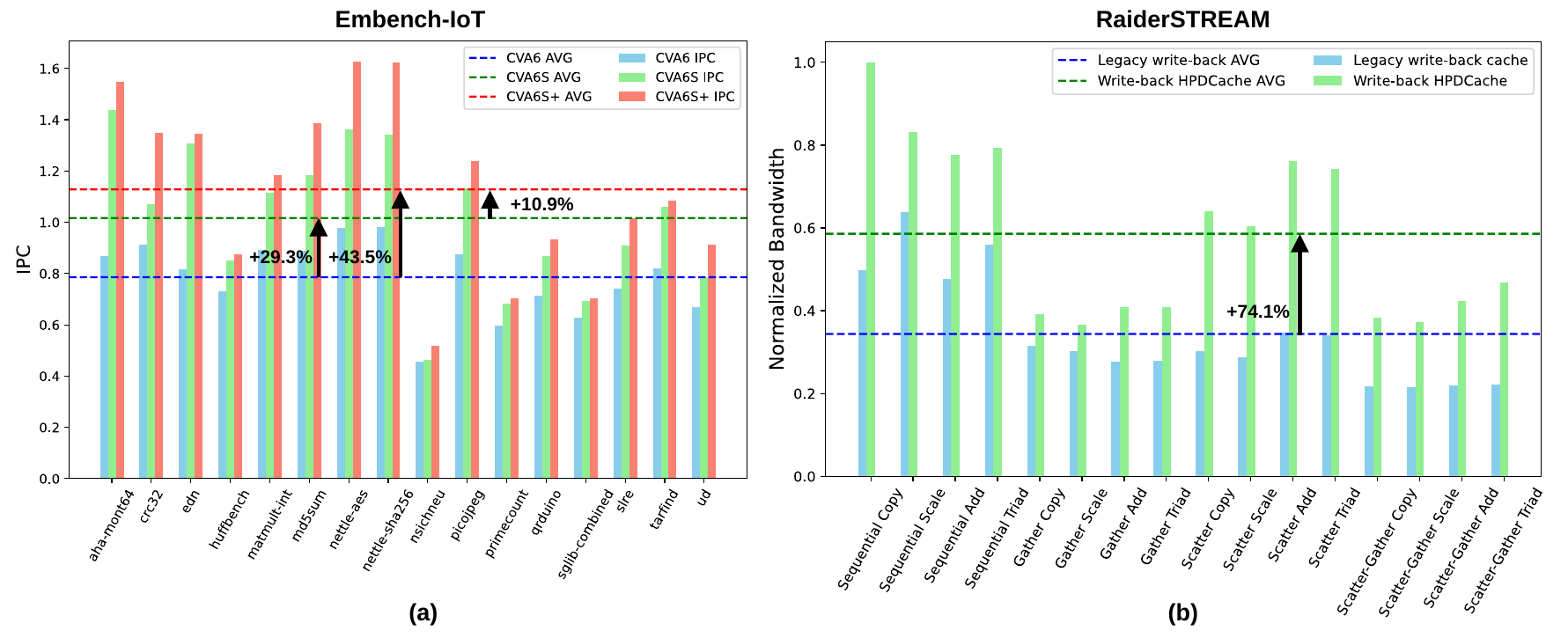}
    \caption{Performance assessment of the different CVA6 versions and cache subsystems. (a) compares IPC improvements on the integer kernels of the Embench IoT suite, (b) the bandwidth gain from adopting the HPDCache.}
    \label{fig:perf}
\end{figure*}

\section{Evaluation}

We use Cheshire \cite{ottaviano2023cheshire} as our testbed, a modular, Linux-capable SoC platform designed to support application-class cores. The SoC is implemented on the Genesys 2 Kintex-7 FPGA. Our evaluation considers the 32-bit versions of CVA6, CVA6S, and CVA6S+, all supporting the \texttt{RV32IMAC} ISA along with the \texttt{a}, \texttt{b}, \texttt{c}, and \texttt{s} bitmanip extensions. The FPU is available only in CVA6 and CVA6S+. Thus, the \texttt{F} extension is not present in CVA6S. The write-back data cache (legacy or HPDCache) is 32 kiB with 8-way associativity, while the instruction cache is 16 kiB with 4-way associativity. An optional Load Unit output register is enabled to improve timing closure, adding one cycle of latency for load operations. To assess pipeline improvements, we use the integer kernels from the Embench-IoT suite, as the working sets fit into the core caches. As shown in Figure \ref{fig:perf}a, we observe an average IPC gain of 43.5\% over the scalar version and a significant 10.9\% improvement over CVA6S. CVA6, CVA6S and CVA6S+ achieve 2.83, 3.41 (+20.2\%) and 3.69 (+30.2\%) CoreMark/MHz respectively.

We use the RaiderSTREAM \cite{beebe2022raiderstream} benchmarks to evaluate the cache subsystem of CVA6S+ with the legacy data cache and the HPDCache. The working set is set to $\times2$ the data cache size. Figure \ref{fig:perf}b reports the bandwidth for sequential and non-sequential memory access patterns, with an average improvement of 74.1\%.

Table \ref{tab:synth-results} presents the topographical synthesis results using the GF22 FDX process under the worst-case conditions of 0.72V at -40°C. Since CVA6S lacks the FPU, resulting in a smaller area, the analysis focuses on CVA6, CVA6S+, and their cache subsystems. The overall area of CVA6S+ increases by only 9.30\%. Excluding the caches, the pipeline area overhead is 28.6\%. Switching to the HPDCache in CVA6S+ leads to an area reduction of 19\% of the DCache due to more efficient SRAM organization. Less than 0.5\% frequency degradation is reported from CVA6 to CVA6S+. HPDcache's area savings enable more synthesis optimizations, boosting the maximum frequency by 13.5\%.

\begin{table}[t]
\centering
\caption{Area at 900 MHz and frequency estimations for the different cores and cache subsystems (16 kiB 4 way ICache, 32 kiB 8 way DCache).}
\label{tab:synth-results}
\resizebox{\columnwidth}{!}{%
\begin{tabular}{cccccc}
\hline
\textbf{Core} & \textbf{DCache} & \textbf{\begin{tabular}[c]{@{}c@{}}Pipeline\\ {[}mm2{]}\end{tabular}} & \textbf{\begin{tabular}[c]{@{}c@{}}ICache\\ {[}mm2{]}\end{tabular}} & \textbf{\begin{tabular}[c]{@{}c@{}}DCache\\ {[}mm2{]}\end{tabular}} & \textbf{\begin{tabular}[c]{@{}c@{}}Max. Freq.\\ {[}MHz{]}\end{tabular}} \\ \hline
CVA6 & HPDCache & 0.057 & 0.041 & 0.077 & 1090 \\
CVA6S+ & Legacy & 0.070 & 0.041 & 0.095 & 960 \\
CVA6S+ & HPDCache & 0.073 & 0.041 & 0.077 & 1095 \\ \hline
\end{tabular}%
}
\end{table}
\section{Conclusion}
We introduce CVA6S+, an enhanced superscalar extension of the CVA6 RISC-V application-class core, adding key features upon its existing dual-issue architecture CVA6S. CVA6S+ integrates the OpenHW HPDCache, combining pipeline improvements with a non-blocking cache subsystem. Our results show a 43.5\% IPC gain and a 74.1\% bandwidth boost with the HPDCache, while incurring under 10\% area overhead.


\bibliography{bibliography} 


\end{document}